\documentclass[]{aa} 
\usepackage{graphicx}
\usepackage{natbib}
\usepackage{txfonts}
\begin{document}

\title{A KOSMA 7 deg$^{2}$ $^{13}$CO 2--1 \& $^{12}$CO 3--2 survey of
  the Perseus cloud}

   \subtitle{I. Structure Analysis}

   \author{K. Sun\inst{1}, C. Kramer\inst{1}, V. Ossenkopf\inst{1,2}, F. Bensch\inst{3}, J. Stutzki\inst{1} \and M. Miller\inst{1}
          }

   \offprints{kefeng@ph1.uni-koeln.de}

   \institute{KOSMA  I. Physikalisches Institut, Universit\"at zu K\"oln, Z\"ulpicher Stra\ss{}e 77, 50937 K\"oln, Germany
         \and
             SRON National Institut for Space Research, P. O. Box 800, 9700, AV Groningen, the Netherlands
         \and
             Radioastronomisches Institut der Universit\"at Bonn, Auf dem H\"ugel 71, 53121 Bonn, Germany
             }

   \date{Received ; accepted }


  \abstract
   {Characterizing the spatial and velocity structure
   of molecular clouds is a first step towards a better understanding
   of interstellar turbulence and its link to star formation. }
   {We present observations and structure analysis results for
   a large-scale ($\sim$ 7.10 deg$^{2}$) $^{13}$CO J = 2--1 and
   $^{12}$CO J = 3--2 survey towards the nearby Perseus molecular cloud
   observed with the KOSMA 3m telescope.}
   {We study the spatial structure of line-integrated and velocity
   channel maps, measuring the $\Delta$-variance as a function of
   size scale. We determine the spectral index $\beta$ of the corresponding
   power spectrum and study its variation across the cloud and across the lines.}
   {We find that the spectra of all CO line-integrated maps of the whole
   complex show the same index, $\beta \approx 3.1$, for scales between about 0.2 and 3\,pc,
   independent of isotopomer and rotational transition. A complementary 2MASS map of
   optical extinction shows a noticeably smaller index of $2.6$. In contrast to the
   overall region, the CO maps of individual subregions show a significant variation
   of $\beta$. The $^{12}$CO 3--2 data provide e.g. a spread of indices between
   2.9 in L\,1455 and $3.5$ in NGC\,1333. In general, active star forming
   regions show a larger power-law exponent. We find that the $\Delta$-variance
   spectra of individual velocity channel maps are very sensitive to optical depth
   effects clearly indicating self-absorption in the densest regions. When studying
   the dependence of the channel-map spectra as a function of the velocity channel
   width, the expected systematic increase of the spectral index with channel width
   is only detected in the blue line wings. This could be explained by a filamentary,
   pillar-like structure which is left at low velocities while the overall molecular
   gas is swept up by a supernova shock wave.}
   {}


   \authorrunning{Sun et al.}
   \titlerunning{A KOSMA 7 deg$^{2}$ CO 2--1 \& 3--2 survey of the Perseus cloud}

   \maketitle
%

\section{Introduction}
The structure of the interstellar medium (ISM) is random to a
large degree, with a complex spatial density distribution and
velocity field. This is evident in large-scale surveys of spectral
line tracers, such as CO, made for parts of the Galactic plane
\citep[e.g.][]{heyer1998,simon2001} and individual molecular
clouds \citep[e.g.][]{bally1987,ut87}. The spatial structure of
the emission has been quantified in terms of its power spectrum
\citep{s87,sbhoz98,e99}. When fitting the azimuthally averaged
power spectrum with a power law, the slope of the power law
$\beta$ provides information on the relative amount of structure
at the linear scales resolved in the image. A pure power law is
expected for structure on the linear scales of a self-similar
turbulent energy cascade. Deviations from a power law are expected
at scales where physical processes insert energy in the turbulence
cascade (outflows, supernovae, super-bubbles, galactic rotation)
and at scales of turbulence dissipation \citep{es04}. Thus, a
study of the scaling behavior of the cloud structure and the
velocity field as traced by the power spectrum of observed
spectral line maps can help to constrain the turbulent energy
cascade in the ISM. A number of power-spectrum studies have been
carried out for the atomic medium using \ion{H}{i} observations
\citep{cd83,g93,ssdss99,dmsgg01,eks01}. They find spectral indices
between about 2.7 and 3.7 for the line-integrated maps
\citep{fhl2004}.

The $\Delta$-variance is an alternative method to determine the
index of the power spectrum of isotropic images \citep{sbhoz98}.
In contrast to the power spectrum, the $\Delta$-variance can be
computed {\it in the spatial domain}. It allows for a better
separation of the intrinsic cloud structure from contributions
resulting from the finite signal-to-noise in the data and the
telescope beam. In addition, problems related to the discrete
sampling of the data can be avoided. \citet{bso01} presented a
$\Delta$-variance analysis for CO maps of the Polaris Flare and
six other molecular clouds. The index $\beta$ is close to 3 for
most clouds, but it steepens in the Polaris Flare from 2.5 to 3.3
for maps with a linear resolution increasing from $\ga 1$\,pc to
$\la 0.1$\,pc. \citet{okh01} used the $\Delta$-variance to study
numerical models of self-gravitating, supersonically turbulent
clouds and compared the results to observations. They showed that
the $\Delta$-variance traces deviations from an inertial scaling
behavior at the scales of driving and dissipation \citep[see
also][]{om02}.

Constraints on the velocity structure of the ISM can be obtained
from an analysis of individual channel maps of a spectral line
cube, and by comparing the results with those obtained for the
line-integrated maps.  \citet{sl01} have analyzed \ion{H}{i}
observations of the Small Magellanic Cloud in this way, noting a
systematic variation of the measured index $\beta$ with the width
of the velocity-channels, thus confirming theoretical predictions
by \citet{lp00}. Using the Boston University/Five College Radio
Astronomy Observatory (BU/FCRAO) Galactic ring survey,
\citet{bs01} found that the index of the channel maps is smaller
than that of the line-integrated maps. It showed a significant
variation as a function of the channel velocity. But $\beta$ did
not increase with increasing velocity channel width, in contrast
to the predictions by \citet{lp00}. Apart from this first attempt,
no systematic structure analysis of velocity-width dependent
channel maps of molecular lines has been performed yet.

We present a new CO survey of the Perseus molecular cloud complex
and analyze the cloud structure in the line integrated and the
channel maps. The observations of the $^{12}$CO 3--2 and $^{13}$CO
2--1 transitions were made with the 3m telescope of the K\"olner
Observatorium f\"ur Sub-Millimeter Astronomie (KOSMA) and cover
the entire Perseus molecular cloud complex (7.1 square degrees).
The Perseus cloud is one of the best examples of a nearby
star-forming region. The cloud is related to the Perseus OB\,2
association \citep{bc86,ut87}, which is located in the area mostly
free of CO emissions northeast of the cloud \citep{ut87}. It
includes a region where intermediate-mass stars form (NGC\,1333),
a young open cluster (IC\,348), and a dozen dense cloud cores with
low levels of star-formation activity (L\,1448, L\,1455, B\,1,
B\,1 EAST, B\,3 and B\,5). 91 protostars and pre-stellar cores
have been identified in a 3 square degree survey of the dust
continuum at 850 and 450 $\mu$m made with the James Clerk Maxwell
Telescope, JCMT \citep{hrfqlc05}.

Imaging observations of the Perseus complex in molecular cloud
tracers exhibit a wealth of substructure, such as cores, shells,
filaments, outflows, jets, and a large-scale velocity gradient
\citep{pbbjn99}. \citet{pbbjn99} compared the structure traced by
$^{13}$CO 1--0 observations to synthetic spectra and find that the
motions in the cloud  must be super-Alfv\'enic, with the exception
of the B1 core, where \cite{gchmt89} and \cite{ctghkm93} detected
a strong magnetic field. \citet{pad03a} find that the structure
function of the line-integrated $^{13}$CO 1--0 map follows a power
law for linear scales between 0.3--3\,pc, and \citet{pad03b}
compared the velocity structure of Perseus to MHD simulations.

A complete census of the stellar content of nearby ($\la 350$\,pc)
molecular clouds (Perseus, Serpens, Ophiuchus, Camaeleon, and
Lupus) is currently obtained by the Spitzer legacy project ``Cores
to Disks'' \citep[c2d,][]{eab03}. Large-scale $^{12}$CO, $^{13}$CO
1--0 and A$_{v}$ maps of the northern clouds were recently
obtained by the COMPLETE team \citep{goodman04}. The KOSMA survey
of Perseus in higher CO transitions traces the warmer and denser
gas due to the elevated critical densities and excitation energies
($\sim$ 10$^{5}$ cm$^{-3}$ and 33.2\,K for CO 3--2) relative to
the J = 1--0 transition. Moreover, $^{12}$CO is largely optically
thick, while $^{13}$CO, being a factor $\sim$ 65 less abundant
\citep{lp90}, is often optically thin, thus tracing column
densities. We plan to extend this work to the other nearby clouds
using the KOSMA and
NANTEN\,2\footnote{http://www.ph1.uni-koeln.de/nanten2}
observatories.

The KOSMA observations are described in Sect.\ 2. The general
properties of the CO data sets are discussed in Sect.\ 3. Sect.\ 4
presents the results of the $\Delta$-variance analysis. The
discussion of the results and a summary are given in Section 5 and
6, respectively.

\section{Observations}
\label{sec-obs}

We mapped the Perseus region simultaneously in $^{13}$CO 2--1 and
$^{12}$CO 3--2 using the KOSMA 3m submillimeter telescope on
Gornergrat, Switzerland, equipped with a dual-channel SIS receiver
\citep{graf98} and acousto optical spectrometers
\citep{schieder89}. Main beam efficiencies and half power
beamwidths (HPBWs) are 68\%, $130''$ at 220\,GHz and 70\%, $82''$
at 345\,GHz. The HPBWs correspond to linear resolutions of
0.22\,pc and 0.14\,pc, where we adopted a distance of 350\,pc
\citep{bb64,hj83,bc86}. All temperatures quoted in this paper are
given on the main beam temperature scale.

For the observations, we divided the $\sim$7.10 deg$^{2}$ region
in $10'\times10'$ fields. Each field was mapped using the
position-switched on-the-fly (OTF) mode \citep{kbssw99,bkds00}
with a sampling of 30\arcsec. Three emission-free off positions
were selected from the $^{13}$CO 1--0 FCRAO map. The pointing was
accurate to within $10''$. The accuracy of the absolute intensity
calibration is better than 15\%, determined with frequent
observations of reference sources. The channel spacing $\Delta
v_\mathrm{ch}$ and the corresponding average baseline noise rms of
the spectra is 0.22\,km\,s$^{-1}$, 0.48\,K for $^{13}$CO 2--1 and
0.29\,km\,s$^{-1}$, 1.02\,K for $^{12}$CO 3--2. The observations
were taken from February to December 2004.

\section{Data Sets}

\subsection{Integrated intensity maps}

\begin{figure*}
   \centering
   \resizebox{\hsize}{!}{\includegraphics[width=\linewidth,angle=-90]{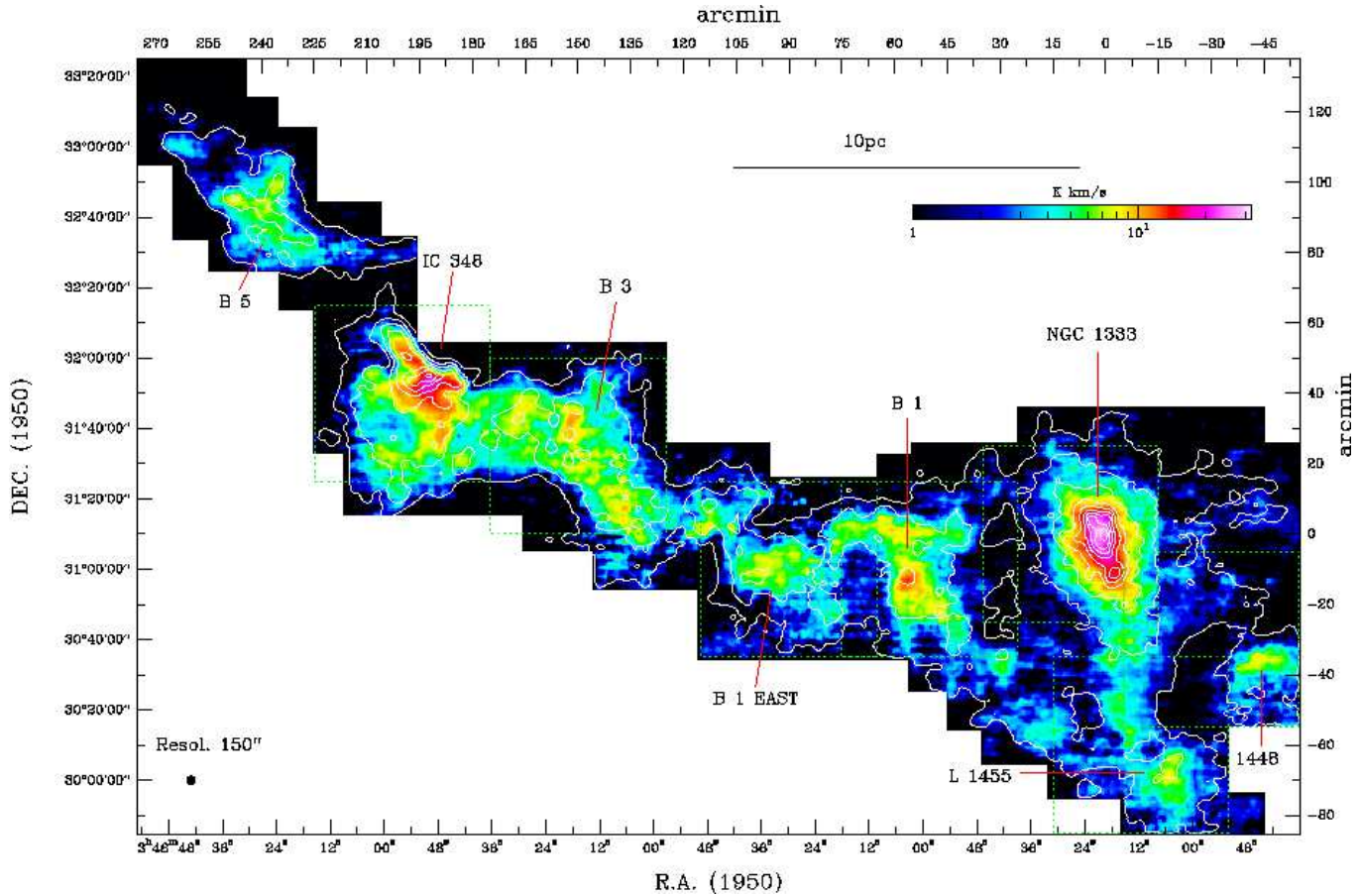}}
\caption{
  The Perseus molecular cloud complex. KOSMA maps of integrated
  intensities of $^{13}$CO 2--1 (colors) and $^{12}$CO 3--2
  (contours) at $150''$ resolution. The integration interval is
  0--16\,km\,s$^{-1}$. Colors run from
  1\,Kkms$^{-1}$ ($\sim$ 1\,$\sigma$) to 32\,Kkms$^{-1}$.  Contours range
  from 6.6\,Kkms$^{-1}$ ($\sim$ 3\,$\sigma$) to 83\,Kkms$^{-1}$ in
  steps of 9\,Kkms$^{-1}$.  The (0,0) position corresponds to
  RA=03:26:00, DEC=$+31$:10:00 (B1950).  Seven sub-regions are marked
  by dotted square boxes of $50'\times50'$.} \label{inten12CO}
\end{figure*}

\begin{figure*}
   \centering
   \resizebox{\hsize}{!}{\includegraphics[width=\linewidth,angle=-90]{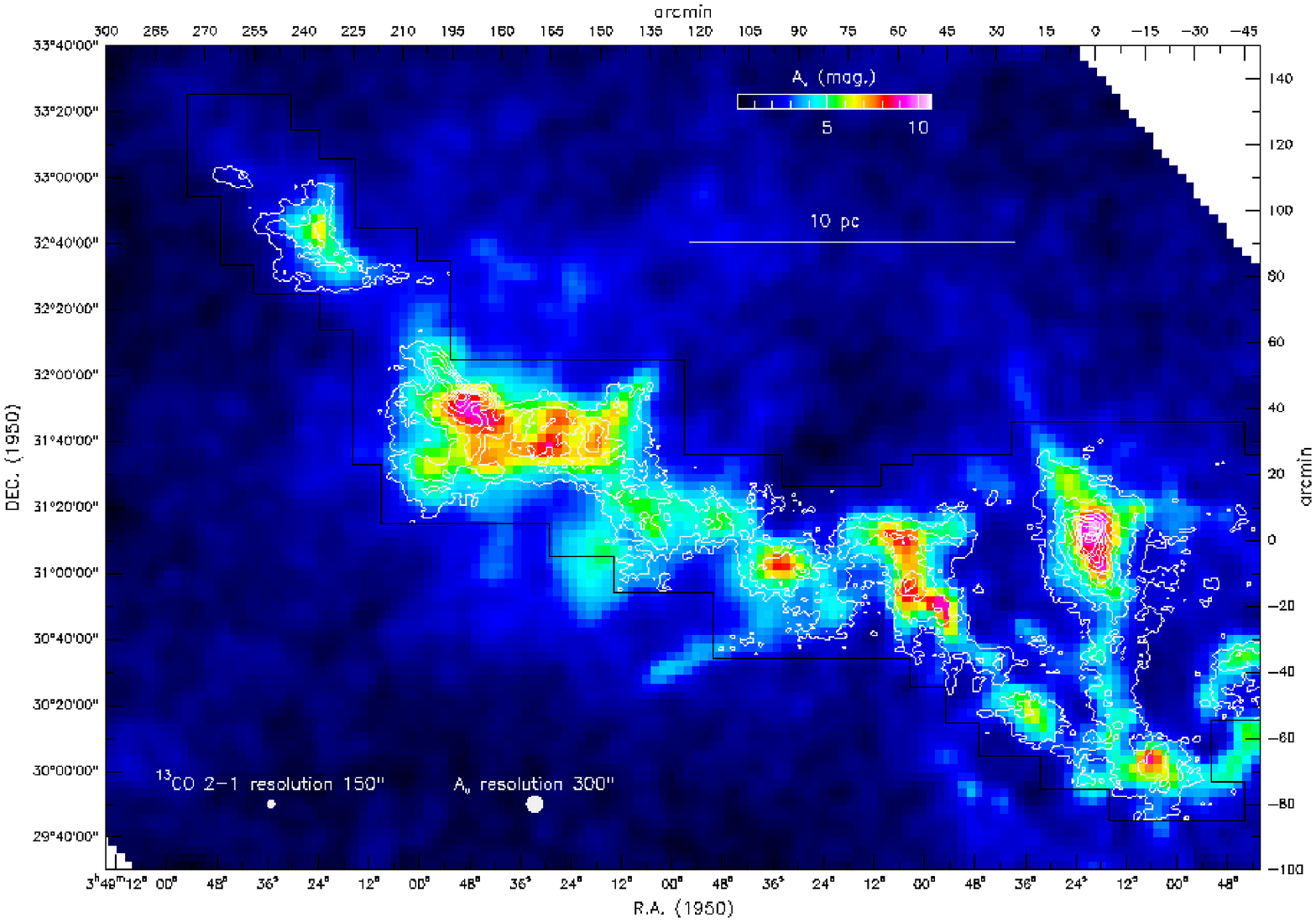}}
   \caption{
     Overlay of $^{13}$CO 2--1 integrated intensities (contours) with
     a map of optical extinctions in colors
     \citep{goodman04,Alves05}. Contours range from 2.7\,Kkms$^{-1}$
     (3\,$\sigma$) to 32\,Kkms$^{-1}$ by 3\,Kkms$^{-1}$. Colors
     range from A$_{v}$ = 1\,mag to 11\,mag.  Resolutions are $2.5'$ for
     $^{13}$CO and $5'$ for $A_\mathrm{V}$. A polygon marks the
     boundary of the $^{13}$CO map. }
     \label{Av13CO}
    \end{figure*}

The maps of velocity integrated $^{13}$CO 2--1 and $^{12}$CO 3--2
emission (Figs.~\ref{inten12CO},\ref{Av13CO}) show the Perseus
region, viz. the well known string of molecular clouds running
over $\sim30$\,pc projected distance from NGC\,1333 and L\,1455 in
the west to B\,1, B\,1 East, and B\,3 in the center, and to
IC\,348 and B ,5 in the east \citep[cf.][]{bc86,ut87}. Generally,
there is a good correlation between $^{12}$CO 3--2 and $^{13}$CO
2--1 integrated intensities.

In the next section, we compare the statistical properties of the
structure seen in the entire Perseus map with the structure seen
in individual regions.  For this, we defined seven boxes of
$50'\times50'$ which roughly coincide with the known molecular
clouds (cf.\,Fig.\,\ref{inten12CO}).

Figure~\ref{Av13CO} shows an overlay of integrated $^{13}$CO 2--1
intensities and a map of optical extinctions
\citep{goodman04,Alves05}, at $2.5'$ and $5'$ resolution,
respectively.  The $^{13}$CO map covers all regions above 7\,mag
and $\sim70$\% of the regions above 3\,mag. A linear least squares
fit to a plot of A$_{v}$ vs. $^{13}$CO 2--1 results in a
correlation coefficient of 0.76. The region mapped in $^{13}$CO
has a mass of 1.7 $\times$ 10$^{4}$ M$_{\odot}$ using the
$A_\mathrm{V}$ data and the canonical conversion factor
[H$_{2}$]/[$A_\mathrm{V}$] = 9.36 $\times$ 10$^{20}$
cm$^{-2}$\,mag$^{-1}$ \citep{bsd78}.

\subsection{Velocity structure}

\begin{figure*}
   \centering
\resizebox{\hsize}{!}{\includegraphics[width=\linewidth]{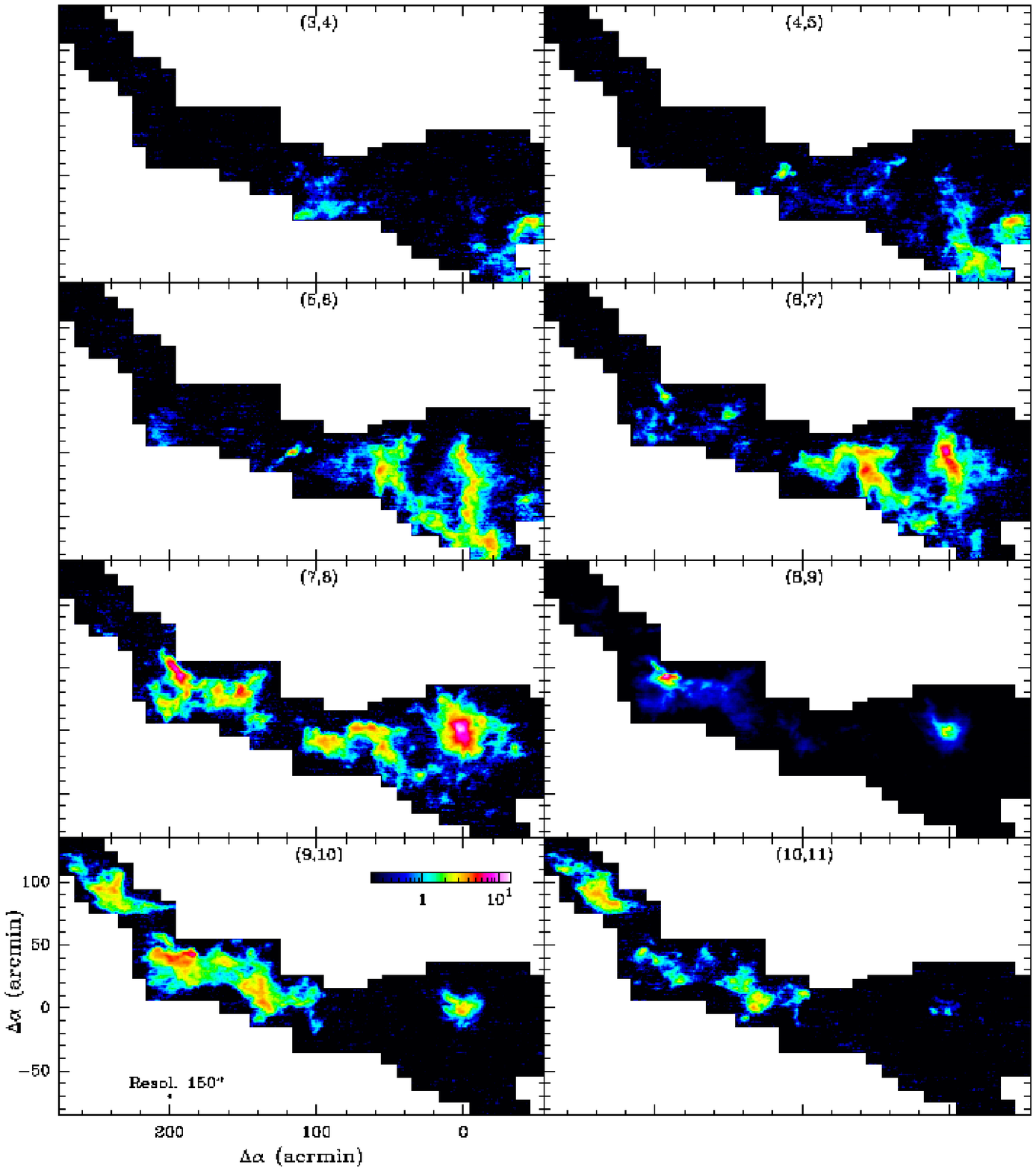}}
\caption{
  $^{13}$CO 2--1 velocity channel maps of the Perseus region. The
  velocity range runs from 3\,km\,s$^{-1}$ to 11\,km\,s$^{-1}$ with an
  interval of 1\,km\,s$^{-1}$ which is indicated on the top of each
  plot. The intensities are plotted from 0.7\,Kkm\,s$^{-1}$ ($\sim$ 1\,$\sigma$) to 15\,Kkm\,s$^{-1}$.}
\label{velchan}
\end{figure*}

Maps of $^{13}$CO 2--1 emission integrated over small velocity
intervals (Fig.\,\ref{velchan}) illustrate the filamentary
structure of the Perseus clouds. The channel maps show the
well-known velocity gradient between the western sources, e.g.
NGC\,1333 at $\sim$ 7\,km\,s$^{-1}$, and the eastern sources, e.g.
IC\,348 at $\sim$ 9\,km\,s$^{-1}$. The channel map integrated
between 5 and 6\,km\,s$^{-1}$ exhibits two filaments originating
at L\,1455, one runs north to NGC\,1333, the second runs
north-east to B\,1.  We will discuss the structural properties of
individual velocity channel maps in the next sections.

\begin{figure}
   \centering
\includegraphics[width=\linewidth]{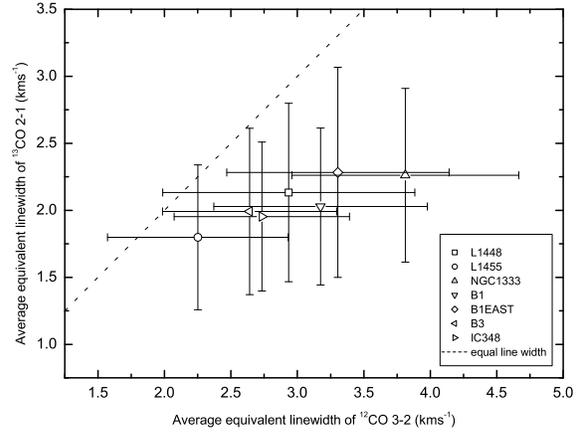}
\caption{Mean and rms of the equivalent line widths $\Delta v_{\rm
    eq}$ of the $^{12}$CO 3--2 and $^{13}$CO 2--1 spectra for the
  observed positions of the seven $50'\times50'$ sub regions
  (Fig.\,\ref{inten12CO}). The dashed line delineates equal widths
  in $^{12}$CO and $^{13}$CO. } \label{linewidth}
\end{figure}

To study the statistics of the velocity field we start with the
distribution of the line widths across the map. Since many spectra
show deviations from a Gaussian line shape, we use the equivalent
line width $\Delta v_{\rm
  eq}=\int T dv / T_{\rm peak}$ as a measure of the line dispersion
along individual lines of sight.  Figure~\ref{linewidth} shows the mean
equivalent line widths and their scatter for the seven sub-regions
shown in Figure~\ref{inten12CO}.

The mean $^{12}$CO widths vary significantly between
2.2\,km\,s$^{-1}$ in the quiescent dark cloud L\,1455 and
3.8\,km\,s$^{-1}$ in the active star forming region NGC\,1333,
while the rms is $\sim0.7$\,km\,s$^{-1}$. In contrast, the
$^{13}$CO widths are smaller and show only a weak trend around
$\sim2$\,kms$^{-1}$.

Several positions in L\,1455, but also in e.g. IC\,348, show small
line widths of $\sim$1\,km\,s$^{-1}$, only a factor of $\sim$8--11
larger than the CO thermal line width, which is $\approx$ 0.16
km\,s$^{-1}$ for a kinetic temperature of 10\,K as was found for
the bulk of the gas in Perseus by \citet{bc86}.

\section{Structure analysis}

In this section, we statistically quantify the spatial structure
observed in the maps, both for the overall structure and for the
structure of individual regions within the Perseus molecular
cloud. We measure the spectral index of the power spectrum using
the $\Delta$-variance analysis, a wavelet convolution technique.
We analyze our new CO data and compare the results with an
equivalent analysis of the FCRAO $^{12}$CO 1--0, $^{13}$CO 1--0
maps and the $A_\mathrm{V}$ Perseus map obtained from 2MASS (Two
Micron All Sky Survey) by the COMPLETE team
\citep{goodman04,Alves05}.

\subsection{The $\Delta$-variance analysis}

The $\Delta$-variance analysis was introduced by \citet{sbhoz98}
as a means to quantify the relative amount of structural variation
at a particular scale in a two-dimensional map or a three-dimensional
data set. The $\Delta$-variance is defined as the variance of an image
$s(\vec{r})$ convolved with a normalized spherically symmetric
wavelet $\odot$ of size $L$
\begin{equation}
\sigma^{2}_\mathrm{\Delta}=\langle \left[s(\vec{r})\ast\odot_\mathrm{L}(\vec{r})\right]^{2} \rangle_{\vec{r}},
\end{equation}
where the asterix denotes the spatial convolution \citep{sbhoz98}.
For structures characterized by a power-law spectrum,
$P(|\vec{k}|)\propto|\vec{k}|^{-\beta}$, the $\Delta$-variance
follows as well a power law, with the exponent $d_\mathrm{\Delta}
= \beta - 2$ in the range $0 \leq \beta \leq 6$.

Unfortunately, the $\Delta$-variance spectrum of any observed data set
does not only reflect the spectral index $\beta$ of the astrophysical
structure, but it is also affected by radiometric noise and the finite
telescope beam \citep{bso01}. Both effects change the spectrum at small
scales. When we ignore the small contribution from beam blurring,
we can write the $\Delta$-variance as
\begin{equation}
\sigma_\mathrm{\Delta}^{2}(L) \approx
a_{1}L^\mathrm{\beta-2}+a_{2}L^\mathrm{d_\mathrm{noise}},
\end{equation}
\citep[Eq. (10) from][]{bso01}. For white noise, $\beta_\mathrm{noise}=0$,
so that $d_\mathrm{noise} = -2$. In the KOSMA data we noticed that
the noise does not follow a pure white noise behaviour, but it
is ``colored'' due to artifacts from instrumental drifts,
baseline ripples, OTF stripes etc. This has to be taken into
account when deriving the cloud spectral index $\beta$ from the
$\Delta$-variance spectra.

\begin{figure}
   \centering
\includegraphics[width=\linewidth]{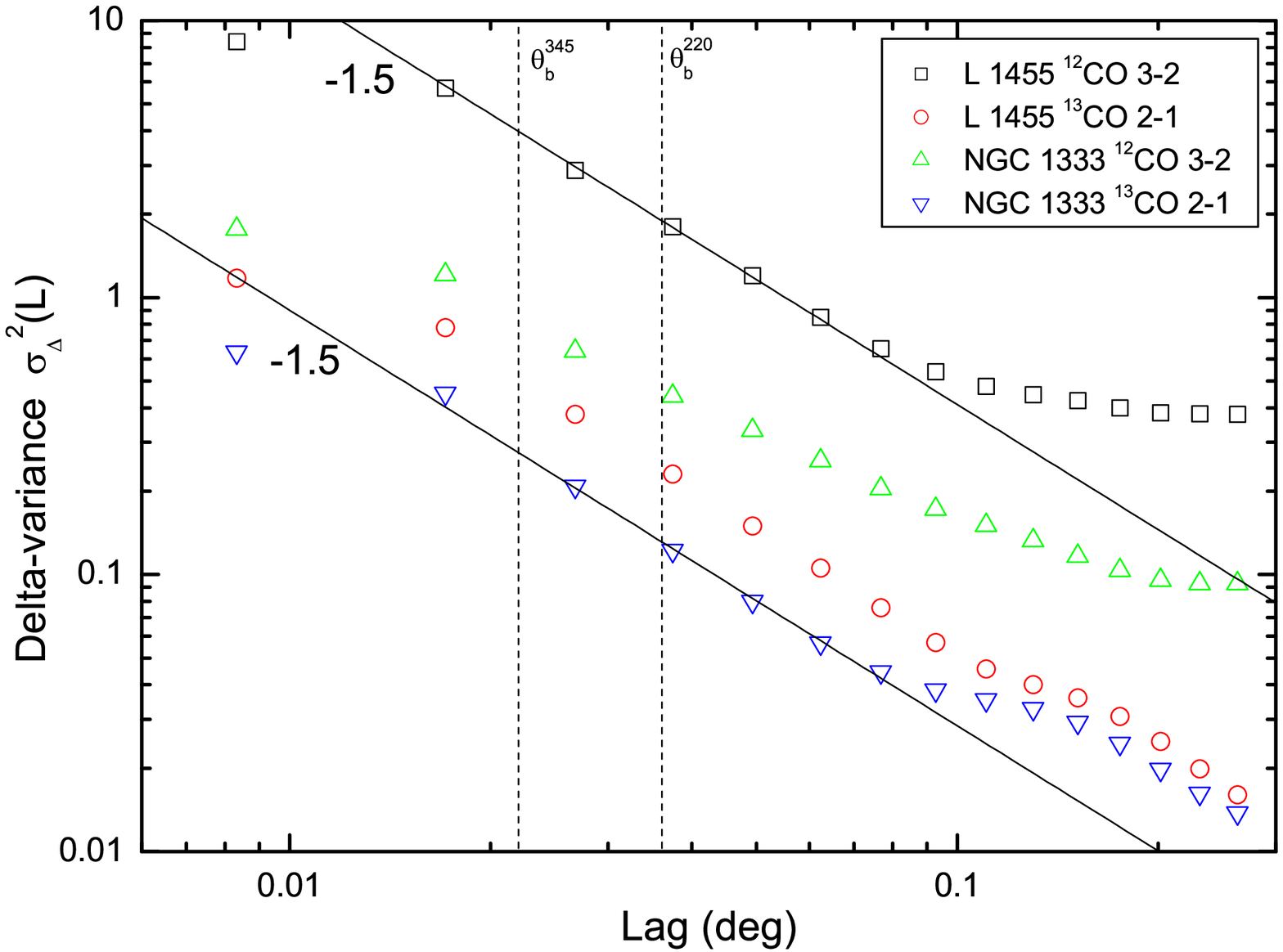}
\includegraphics[width=\linewidth]{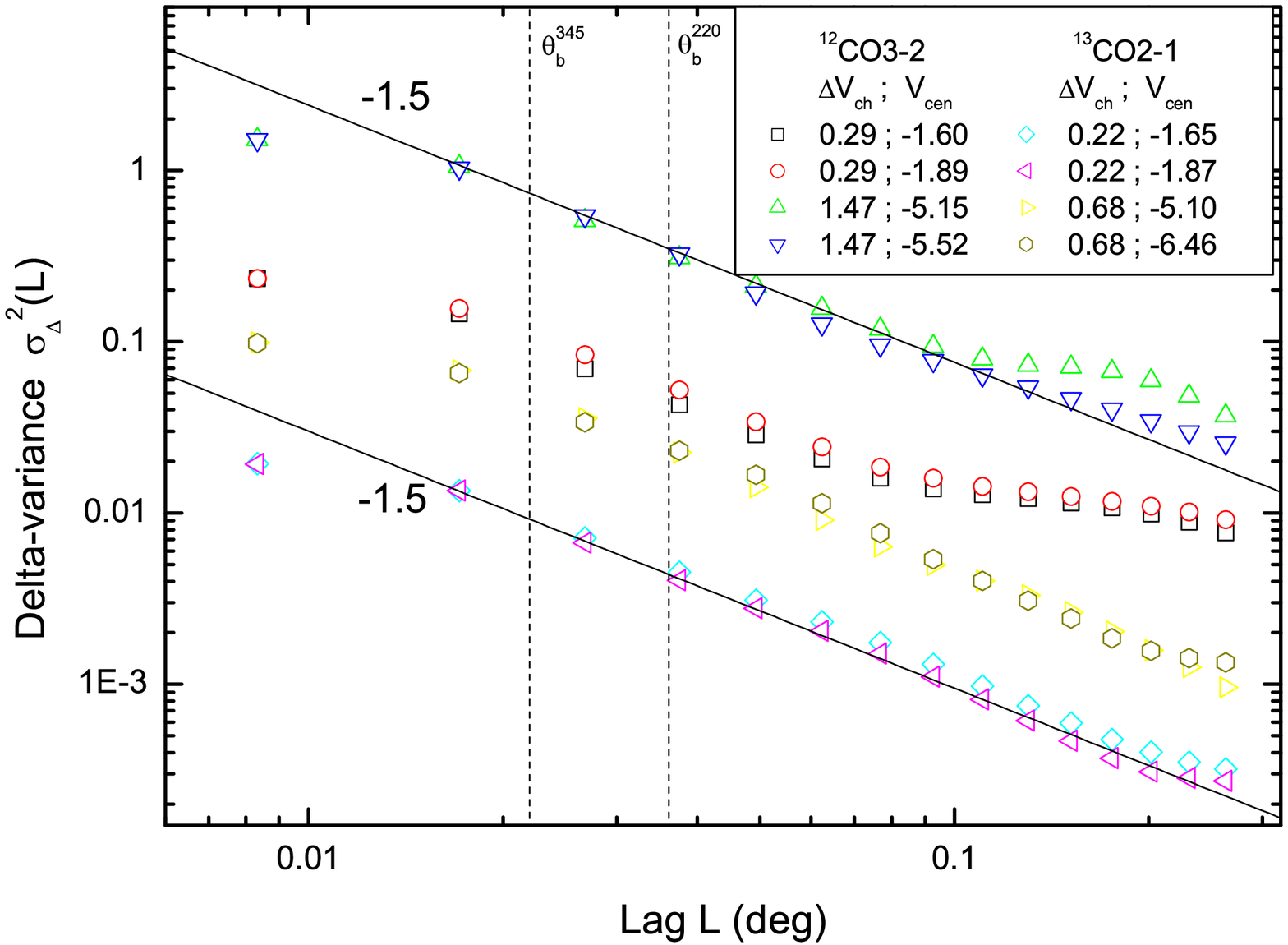}
\caption{$\Delta$-variance analysis of the off-line channel maps.
In the upper plot a velocity span corresponding to the integrated
intensity maps is used. The two regions representing opposite
extremes in the structural behaviour, NGC 1333 and L1455, show
about the same spectral index of the colored noise in both
transitions for small lags. In the lower plot, the influence of
different velocity spans, as used in the velocity channel analysis
(Sect. \ref{sect_vca}), is studied for L1455. The colored noise
index $d_\mathrm{noise}$ is nearly constant independent of
species, transition, velocity range $\Delta$v$_{\rm ch}$, and
center velocity v$_{\rm cen}$.} \label{offline}
\end{figure}

Thus we measured the spectral index of the colored noise
$d_\mathrm{noise}$ by analyzing maps created from velocity
channels which do not see any line emission but which cover the
same velocity width as the actual molecular line maps. The result
is shown in Fig. \ref{offline}. We find a nearly constant index
$d_\mathrm{noise}\approx -1.5$ for all off-line channels at scales
between about 1 and 6\arcmin. At larger lags, the noise deviates
from the $\beta$ = 0.5 behaviour, but this does not affect the
structure analysis as the absolute noise contribution is
negligible there.

For the FCRAO data and the COMPLETE $A_\mathrm{V}$ map we have no
emission-free channels available so that we cannot perform an
equivalent noise fit there. The $\Delta$-variance at small lags
shows however no indications for a deviation from the pure white
noise behaviour, so that we stick to $d_\mathrm{noise}=-2$ for the
fit of these data.

\subsection{Integrated intensity maps}

\begin{figure}
   \centering
\includegraphics[width=\linewidth]{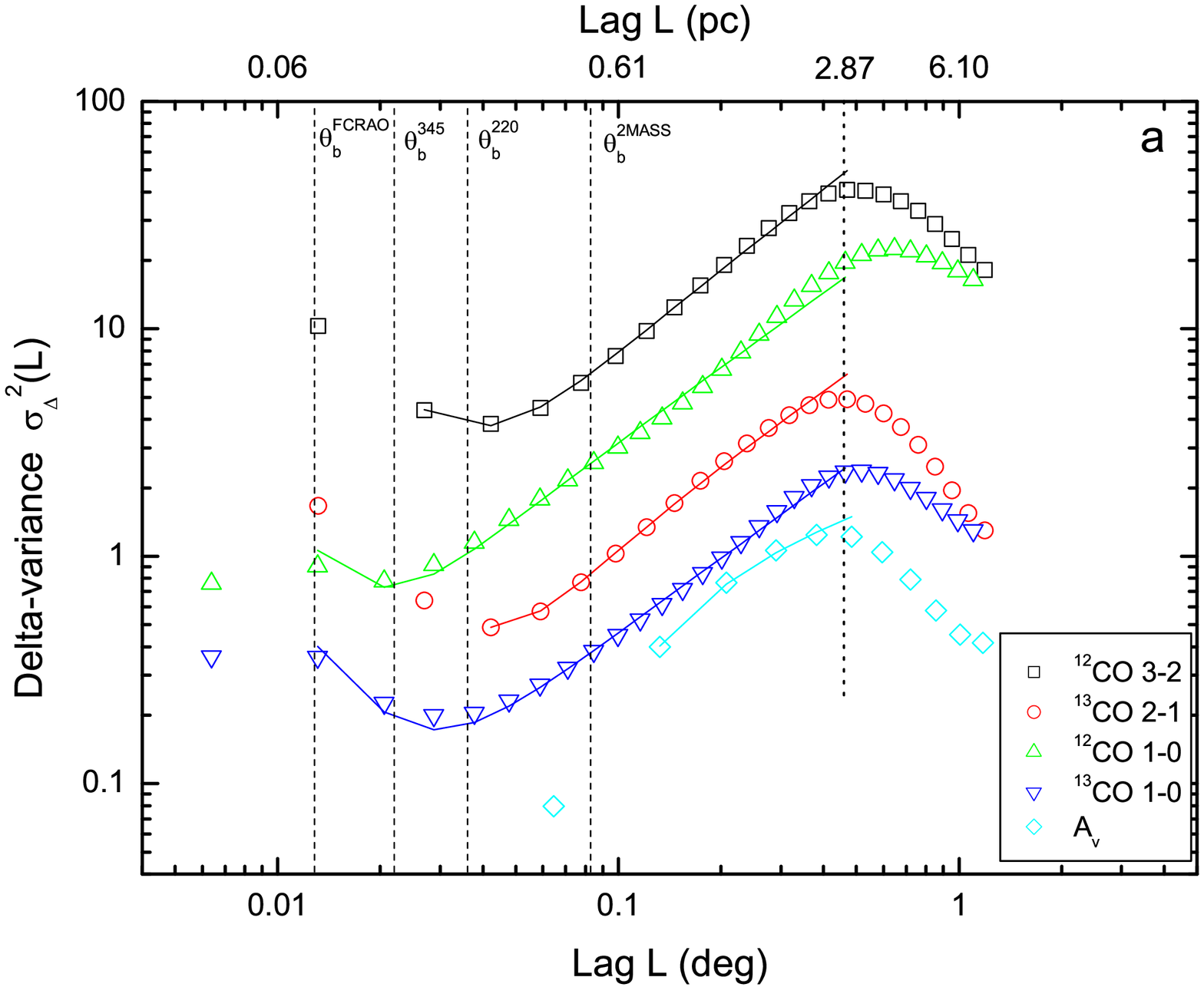}
\includegraphics[width=\linewidth]{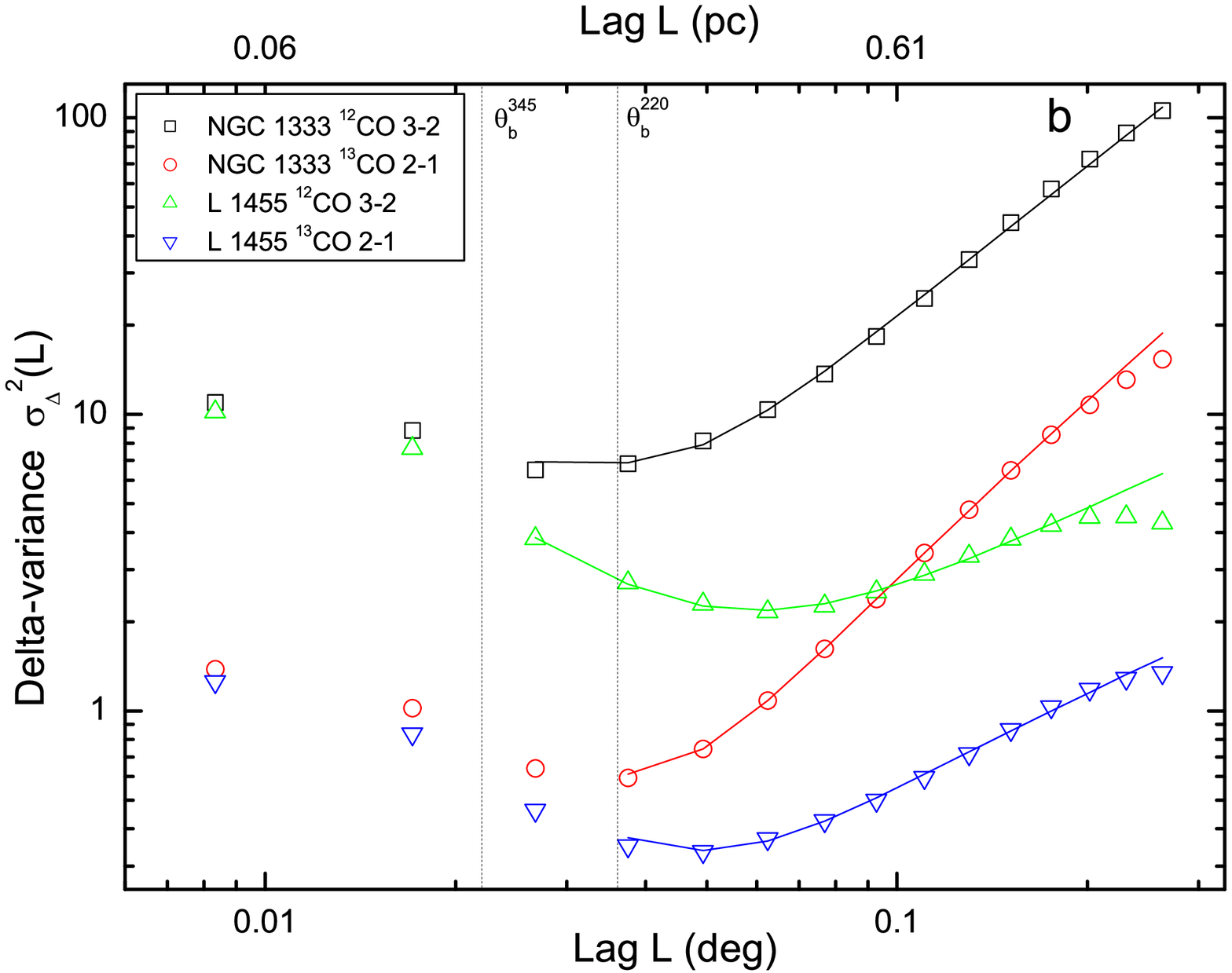}
\caption{ $\Delta$-variance spectra of integrated intensities.
  {\bf a)} Spectra obtained from the CO maps and the $A_\mathrm{V}$
  data of the region mapped with the KOSMA telescope.
  {\bf b)} Spectra of integrated intensity maps of two 50\arcmin $\times$ 50\arcmin sub-regions:
  NGC\,1333 and L\,1455.  Power-law fits to the data corrected for
  noise and beam-blurring are indicated as solid lines.}
\label{deltaint}
\end{figure}

Figure~\ref{deltaint} compares the $\Delta$-variance spectra of
the different integrated intensity maps for the entire region
mapped with KOSMA (see Fig.~\ref{inten12CO})\footnote{Note that
the area covered by the FCRAO is slightly smaller than that
observed with KOSMA.}. When corrected for the observational noise,
the $\Delta$-variance spectra of all maps follow power laws
between the linear resolution of the surveys and about 3\,pc
(Table~\ref{deltatable}). The good agreement of the spectral
indices obtained from the different CO data is remarkable. They
cover only the narrow range between 3.03$\pm$0.14 and
3.15$\pm$0.04. In contrast, the extinction data result in a
significantly lower index. This indicates a more filamentary
structure in $A_\mathrm{V}$. When we actually compare the
$A_\mathrm{V}$ map with $^{13}$CO 2--1 data smoothed to the same
resolution, it is also noticeable by eye that the $A_\mathrm{V}$
map looks more clumpy or filamentary than the $^{13}$CO map. This
indicates that $^{13}$CO does not trace all details of the cloud
structure, but rather measures the more extended, and thus more
smoothly distributed gas.

All $\Delta$-variance spectra show a turnover at about 3\,pc. To
test whether this peak measures the real width of the Perseus
cloud or whether it is produced by the elongated shape of the CO
maps, we have repeated the $\Delta$-variance analysis for the
$A_\mathrm{V}$ data of the entire region shown in
Figure~\ref{Av13CO}. In this case we find almost the same spectrum
below 3~pc, but instead of a turnover only a slight decrease of
the slope at larger lags. Thus we have to conclude that the
$\Delta$-variance spectra of the CO maps at scales beyond 3~pc are
dominated by edge effects, due to the shape of the maps, so that
these scales should be excluded from the analysis. In
Figure~\ref{deltaint} we compare only spectra for the same region,
i.e. the $\Delta$-variance spectra of the $A_\mathrm{V}$ data of
the region also mapped with KOSMA.

As it is not guaranteed that the structure of the overall region
is representative for individual components, we have also applied
the $\Delta$-variance analysis to the KOSMA data of the individual
clouds contained in the seven $50\arcmin \times 50\arcmin$
subregions shown in Fig.~\ref{inten12CO}. The results of the
power-law fits to the $\Delta$-variance spectra are listed in
Table~\ref{deltatable_indi}. They differ significantly between the
individual regions. The active star-forming region NGC\,1333 shows
the highest spectral indices in both transitions. The low end of
the spectral index range is formed by the dark cloud L\,1455
together with the environment of the young cluster IC\,348. The
$\Delta$-variance spectra of the two extreme examples NGC\,1333
and L\,1455 are shown in Fig.~\ref{deltaint}b. Starting from the
same noise values at small scales the spectra of the two regions
show an increasing difference in the relative amount of structure
at large scales reflected by the strongly deviating spectral
indices. Altogether, we find high indices as characteristics of
large condensations for the regions with active star formation and
lower indices quantifying more filamentary structure for dark
clouds, but IC\,348 as an exception to this rule, showing also a
very filamentary structure.

\begin{table}
\caption[]{\label{deltatable} {\small Results of the $\Delta$-variance
analysis of the integrated CO maps and the $A_\mathrm{V}$ data for
the region mapped with KOSMA (Fig.\,\ref{inten12CO}).
}} \centering
\begin{tabular}{ccrcc}
\hline
Transition & Telescope & resol.  & Fit Range & $\beta$\\
   & & [\arcmin] & [\arcmin] &  \\
\hline
$A_\mathrm{V}$ & 2MASS      & 5     & 5.0-28 & 2.55 $\pm$ 0.02 \\
$^{13}$CO 1--0 & FCRAO      & 0.77  & 0.8-28 & 3.09 $\pm$ 0.09 \\
$^{12}$CO 1--0 & FCRAO      & 0.77  & 0.8-28 & 3.08 $\pm$ 0.04 \\
$^{13}$CO 2--1 & KOSMA      & 2.17  & 2.2-28  & 3.03 $\pm$ 0.14 \\
$^{12}$CO 3--2 & KOSMA      & 1.37  & 1.4-28  & 3.15 $\pm$ 0.04\\
\hline

\end{tabular}
\end{table}

\begin{table}
\caption[]{\label{deltatable_indi} {\small Results of the
$\Delta$-variance analysis of the KOSMA data for seven 50\arcmin
$\times$ 50\arcmin sub-regions of the cloud
(Figure~\ref{inten12CO}). The spectral indices $\beta$ were fitted
in the size range 2.2-14$'$ for the $^{13}$CO 2--1 and in the size
range 1.4-14$'$ for the $^{12}$CO 3--2 data.}} \centering
\begin{tabular}{lcc}
\hline
Region & $\beta(^{13}$CO$\, 2-1)$ & $\beta(^{12}$CO$\, 3-2)$ \\
\hline
L\,1448              & 2.96 $\pm$ 0.42 & 3.41 $\pm$ 0.16 \\
L\,1455              & 2.86 $\pm$ 0.09 & 2.85 $\pm$ 0.30 \\
NGC\,1333            & 3.76 $\pm$ 0.48 & 3.52 $\pm$ 0.11 \\
B\,1                 & 3.14 $\pm$ 0.29 & 3.00 $\pm$ 0.20 \\
B\,1 EAST            & 3.16 $\pm$ 0.09 & 3.39 $\pm$ 0.09 \\
B\,3                 & 3.36 $\pm$ 0.09 & 3.14 $\pm$ 0.06 \\
IC\,348              & 2.71 $\pm$ 0.42 & 3.06 $\pm$ 0.24 \\
\hline
\end{tabular}
\end{table}

\subsection{Velocity channel maps}
\label{sect_vca}

When performing the $\Delta$-variance analysis not only for maps
of integrated intensities, but for individual channel maps we
obtain additional information on the velocity structure of the
cloud. In the velocity channel analysis (VCA), introduced by \citet{lp00},
the change of the spectral index of channel maps as a function
of the channel width was used to simultaneously determine the
scaling behavior of the density and the velocity fields from a
single data cube of line data. Here we conduct such a study for
the KOSMA CO data.

\begin{figure}
   \centering
   \includegraphics[width=\linewidth]{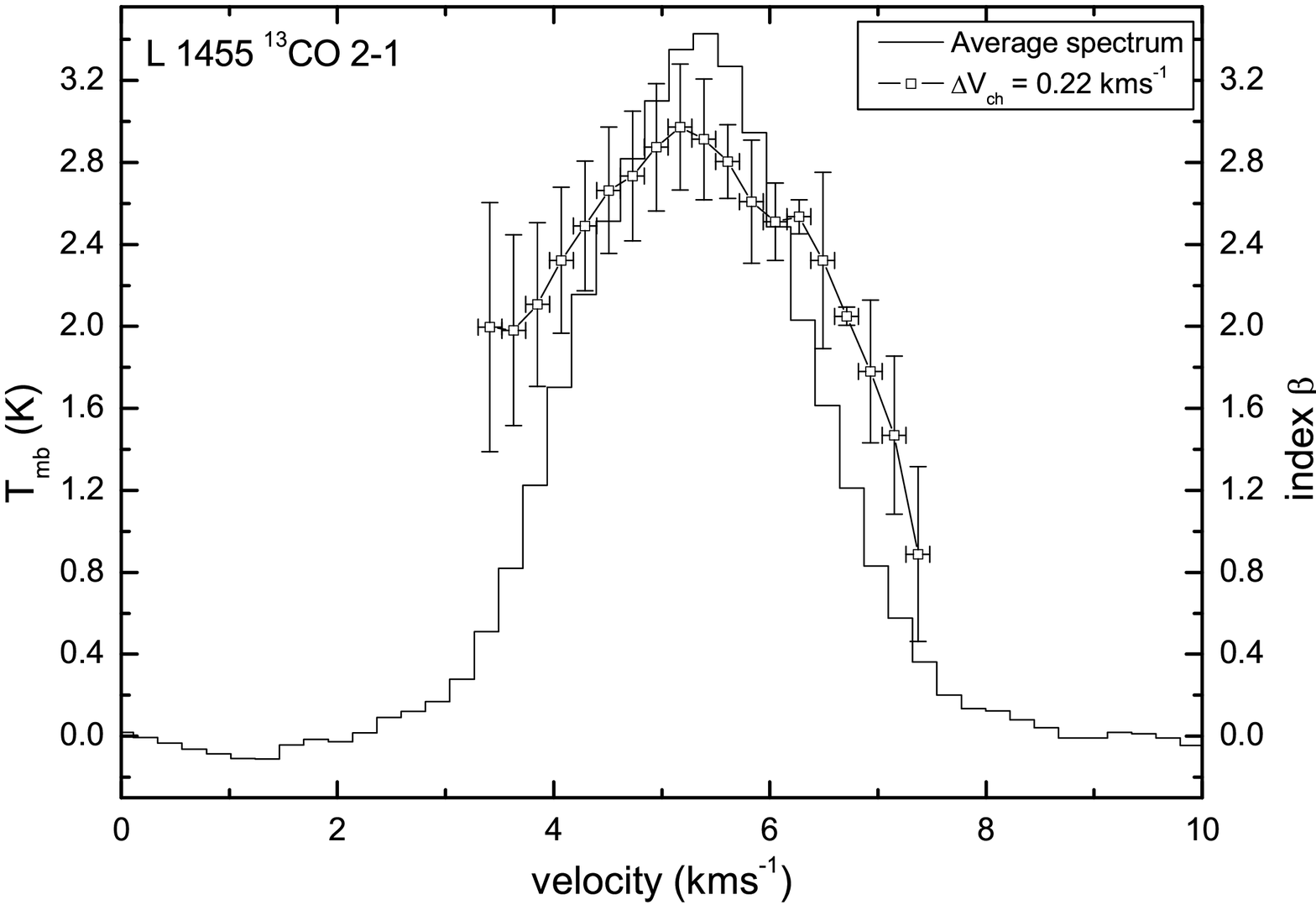}
\caption{Comparison of the index spectrum of the
  $^{13}$CO 2--1 data in L\,1455 with the average line profile.  The
  index spectrum is created by power-law fits to the $\Delta$-variance
  spectrum of individual channel maps ($\Delta$v$_\mathrm{ch}$ =
  0.22\,km\,s$^{-1}$). The vertical error bars represent the
  uncertainty of the fit. The horizontal error bars indicate the
  velocity channel width.}
\label{deltaone}
\end{figure}

We start with the analysis of individual channel maps as they are
provided by the
channel spacing $\Delta$v$_\mathrm{ch}$ of the backends
(cf.\,\S\ref{sec-obs}).
 For all channel maps we perform the $\Delta$-variance analysis and
 fit power laws to the measured structure for all lags between the
 telescope beam size and the maximum scale resolved by the
 $\Delta$-variance (about 1/4 of the map size).
 As a result we get the power-law index as a function of the channel
 velocity, a curve which we call {\it index spectrum}. As an example
 we show the index spectrum obtained for the $^{13}$CO 2--1 data in
 the L\,1455 region in Fig.~\ref{deltaone}.  The spectrum is always
 truncated at velocities where the average line temperature is lower
 than the noise rms.

The overall structure of the index spectrum is similar to the line
profile. The largest spectral indices are found at velocities close
to the average line peak. This may implicate that extended smooth
structure provides the major contribution to the overall emission,
while the velocity tail of this structure is formed by small-scale
features. However, the indices show an asymmetric behavior with
respect to the blue and the red wing. The indices drop steeply
to a noise-dominated value at the red wing, while the blue wing
shows only a very shallow decay. Even at the noise limit, noticeable
structure is detected in the channel maps there.

\begin{figure*}
   \centering
\mbox{
\begin{minipage}[b]{8cm}
\includegraphics[width=8cm]{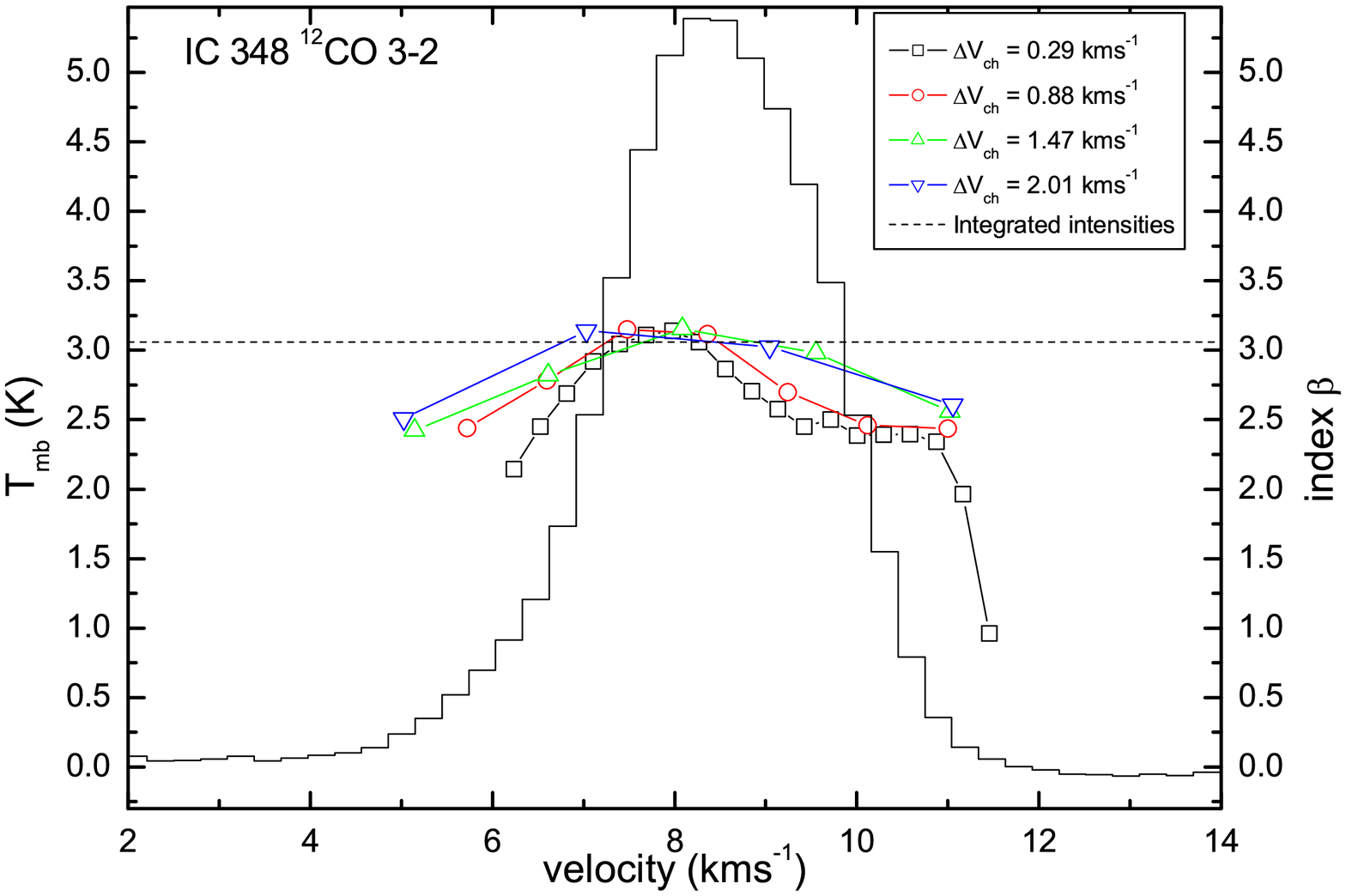}
\end{minipage}
\begin{minipage}[b]{8cm}
\includegraphics[width=8cm]{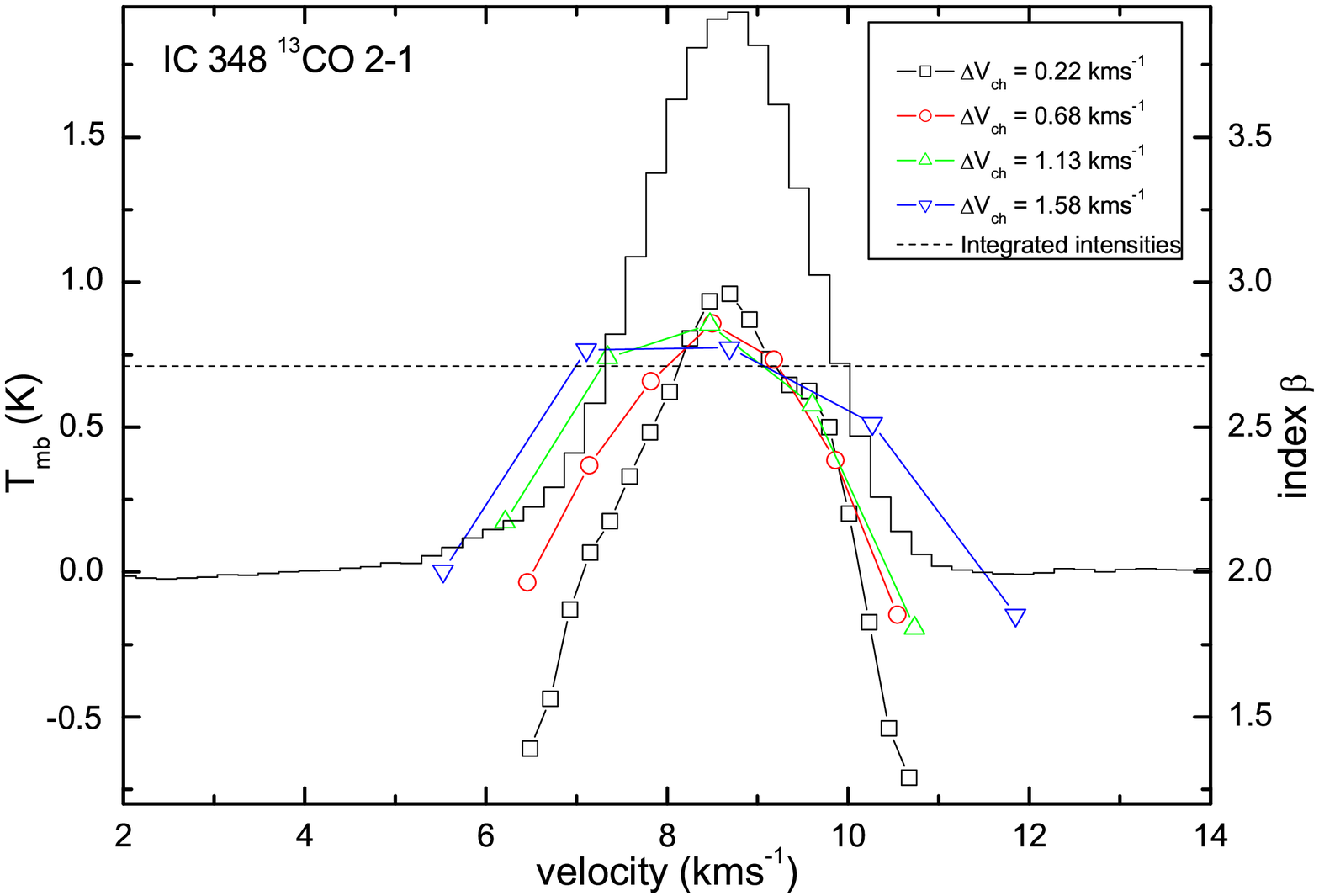}
\end{minipage}}

\mbox{
\begin{minipage}[b]{7.5cm}
\includegraphics[width=7.5cm]{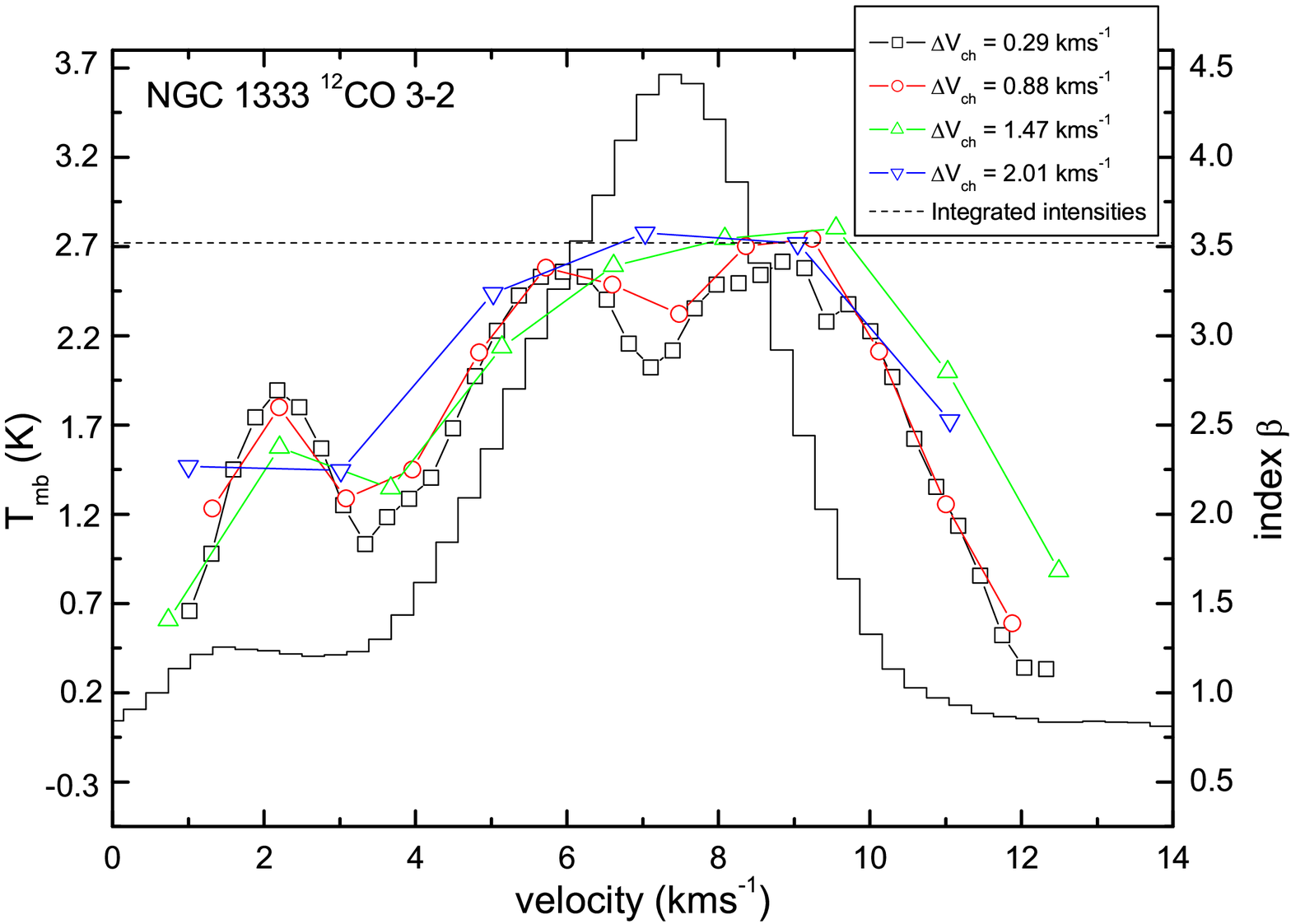}
\end{minipage}
\begin{minipage}[b]{8cm}
\includegraphics[width=8cm]{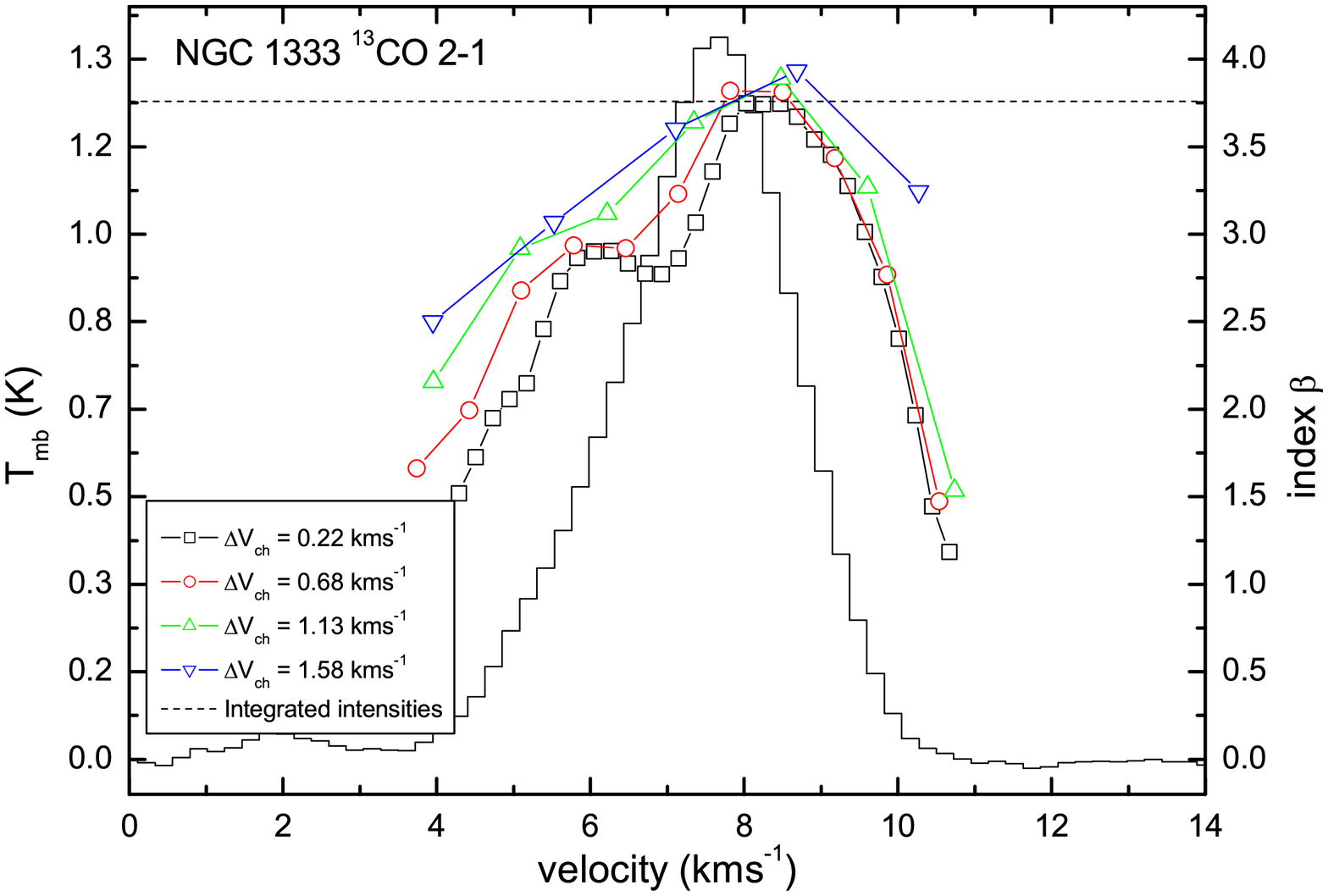}
\end{minipage}}

\mbox{
\begin{minipage}[b]{7.7cm}
\includegraphics[width=7.7cm]{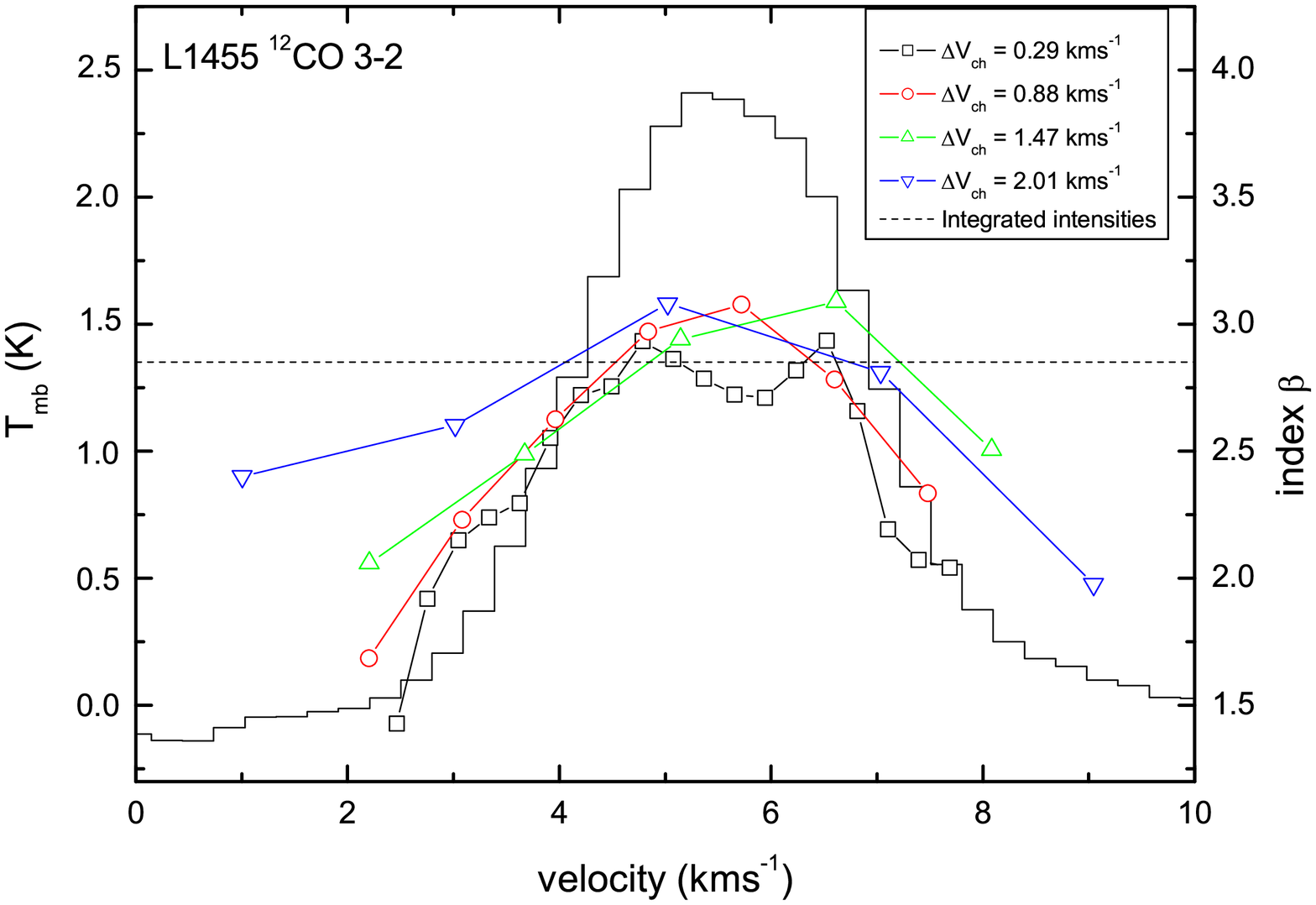}
\end{minipage}
\begin{minipage}[b]{8cm}
\includegraphics[width=8cm]{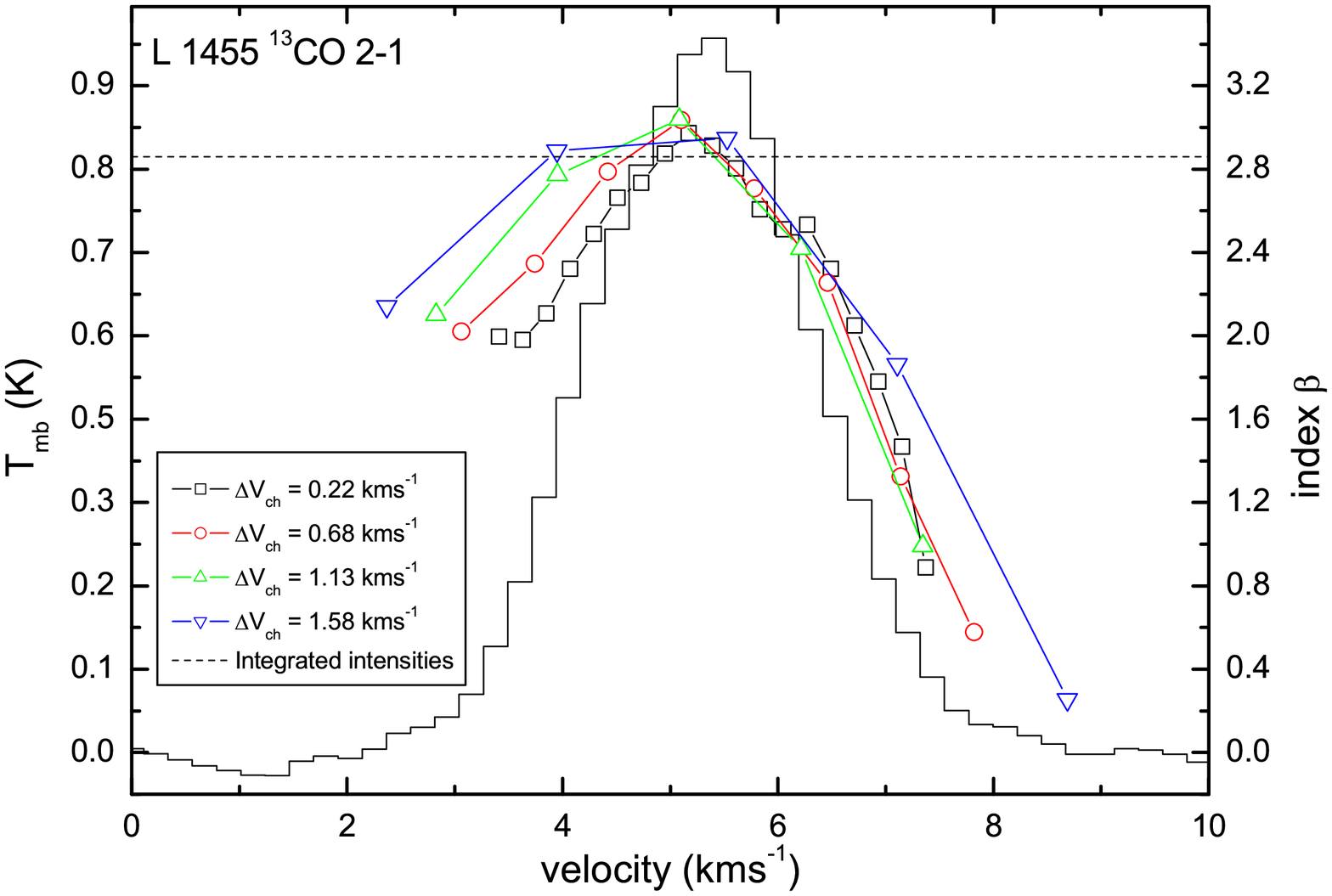}
\end{minipage}}
\caption{Comparison of the index spectra obtained for different
velocity channel widths with the average line profile. The
upper plots show the results for IC\,348, the central plot
NGC\,1333 and the lower plot L\,1455. For the left column we
used the $^{12}$CO 3--2 data, the right column represents the
$^{13}$CO 2--1 data. The different symbols indicate the results
from different velocity channel widths. The dashed lines represents the
index of the integrated intensity maps.}
    \label{deltachan}
    \end{figure*}

For the full velocity channel analysis, the index spectrum has to
be computed for different velocity channel widths \citep{lp00}.
Thus we have binned the data to averages of three, five, and seven
velocity channels and computed the index spectra for these
binned channel maps. In Fig.~\ref{deltachan} we show the
results for three examples: IC\,348, NGC\,1333 and L\,1455. For the
sake of clarity the error bars of the index spectra were omitted
in these plots.

The overall structure of the index spectra is similar to
Fig.~\ref{deltaone} for all sources, transitions and channel
widths. In most cases we find the asymmetry of a shallower blue
wing relative to the red wing. When looking at narrow velocity
channels, we find a dip in the centre of the index spectrum for
the $^{12}$CO 3--2 data of NGC\,1333 and L\,1455. A slight
indication of such a dip is also present in the $^{12}$CO 3--2
data if IC348 and in the $^{13}$CO 2--1 data of NGC\,1333. This is
due to optical depth effects. When we check individual spectra in
those regions, we see self-absorption in several positions. This
leads to a more filamentary appearance of the central channel maps
reflected by this dip in the index spectra. It is interesting to
notice that the VCA is more sensitive to self-absorption than the
average spectrum.

When increasing the channel width by binning, the self-absorption
dip is smoothed out, so that the resulting index spectra peak
again close to the peak velocity of the line temperature. In all
situations where the self absorption is negligible, the indices
for the line core channels are almost independent from the channel
width. The indices for the line integrated intensities always fall
slightly below the peak indices, as they represent an average
which is typically dominated by the line cores.

In the red line wings, most indices remain approximately constant
when increasing the velocity width, except for the largest bin
width where the contribution from the core leads to an observable
increase. In the blue wing, we find a monotonic growth of the
spectral indices with the channel width for both tracers in all
three regions. The additional peak at 2~km\,s$^{-1}$ visible in
the $^{12}$CO 3--2 data of NGC\,1333 stems from a separate dark
cloud which is also contained in the NGC\,1333 map.

\begin{figure}
   \centering
   \includegraphics[width=\linewidth]{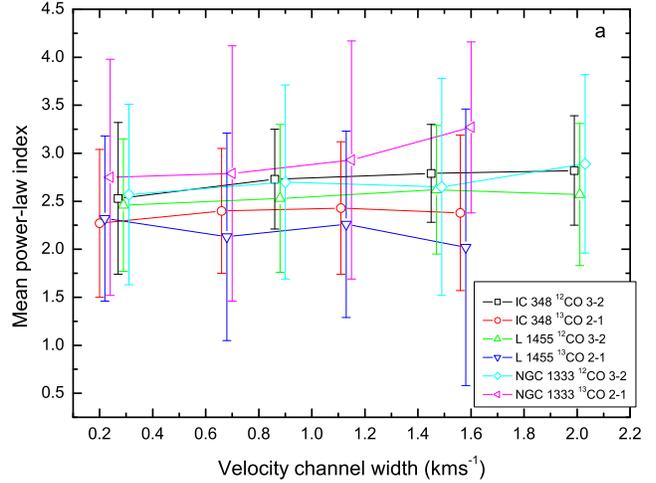}
   \includegraphics[width=\linewidth]{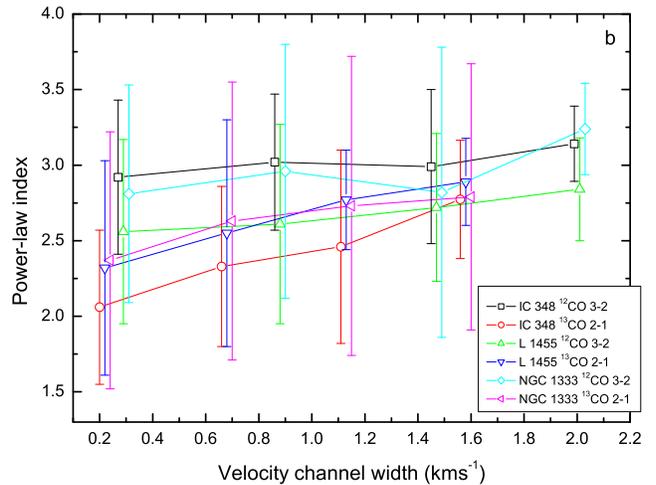}
   \caption{Average spectral indices of the channel maps as a function
   of the channel width. {\bf a)} shows the average over the full line
   width. {\bf b)} represents only the indices in the blue ling wings.
   The line wing components are centered at 5\,km\,s$^{-1}$ for NGC\,1333,
   at 4\,km\,s$^{-1}$ for L\,1455, and at 7\,km\,s$^{-1}$ for IC\,348.
   The error bars contain the fit error and the standard deviation of the
   indices within the considered velocity range. To avoid overlapping
   error bars in the plot, we have shifted the points for IC\,348 and
   NGC\,1333 by $\pm 0.02$~km\,s$^{-1}$ relative to their actual position.  }
\label{slopevel}
\end{figure}

Figure~\ref{slopevel} summarizes the relation between the spectral
indices and the velocity channel width. In Fig.~\ref{slopevel}a we
plot the average spectral index over the line as a function of the
channel width for the six data sets presented in
Fig.\,\ref{deltachan}. Figure ~\ref{slopevel}b contains the
analysis when restricted to a 2~km\,s$^{-1}$ window in the blue
line wings.  The error bars contain the standard deviation of the
index variation across the line and the fit errors. They are
necessarily large because of the systematic variation of indices
over the velocity range. In contrast to similar analysis by
\citet{dmsgg01,sl01} we find no significant systematic variation
of the mean line index as a function of channel width
(Figure~\ref{slopevel}a). In contrast to the average of the index
spectrum we find a continuous increase of the spectral index with
the channel width when restricting the analysis to the blue wing
(Figure~\ref{slopevel}b). The average index steepens from about
2.8 to about 3.1 in the $^{12}$CO 3--2 maps and from about 2.4 to
about 2.8 in the $^{13}$CO 2--1 maps. As discussed above we find
no systematic trend in the red wings. This indicates that the
average spectral index taken over the full line profile provides
no measure for the velocity structure in our CO maps while the
peculiar behaviour in the blue wings needs further investigation.

\section{Discussion}

\subsection{Integrated intensity maps}

Besides the $\Delta$-variance, other tools have been used to
characterize interstellar cloud structure. The second-order
structure function for an observable $s(\vec{r})$ is $S_\mathrm{2}
= \langle|s(\vec{r}) - s(\vec{r}+\delta
\vec{r})|^{2}\rangle_{\vec{r}}$ which is treated as a function of
the absolute value of the increment $|\delta \vec{r}|$
\citep{es04}. \citet{pad03a} computed the structure function of
the integrated intensity map of $^{13}$CO 1--0 in Perseus.  A
power-law fit to S$_{2}$ $\propto$ $\delta$r$^\mathrm{\zeta}$ over
a range of 0.3 to 3\,pc provided an index $\zeta$ of 0.83.  The
index of the structure function is related to the power spectral
index by $\zeta=\beta-2$ for $2<\beta<4$ \citep{sbhoz98} resulting
in $\beta=2.83$. This result of \citet{pad03a} agrees within the
error bars with the indices found by the $\Delta$-variance
analysis of the Perseus maps of integrated CO intensities over
almost the same linear range (see Table~\ref{deltatable}).

However, the $\Delta$-variance spectra of individual regions show
significant variations of the spectral index as discussed above.
For $^{13}$CO 2--1, these span the range between $\beta=2.86$ in
L1455 and 3.76 in NGC\,1333 (Table\,\ref{deltatable_indi}). The
analysis of different sub-sets in molecular cloud complexes thus
provides additional and complementary information on the structure
of the cloud complex.

\subsection{Velocity channel maps}

The velocity channel analysis (VCA), was used previously by
\citet{dmsgg01} to study \ion{H}{i} maps of two regions in the 4th
Galactic quadrant. One of the regions is rich in warm \ion{H}{i}
gas, the other is rich in cool \ion{H}{i} gas. For the warm gas,
\citet{dmsgg01} find a systematic increase of the mean index with
velocity channel width. The cold gas at lower latitudes behaves
differently and shows rather constant indices of 2.7--3.1.

The latter results resemble the outcome of the VCA of the
$^{12}$CO and $^{13}$CO channel maps in Perseus presented above.
We find similar indices for the velocity integrated maps of the
full region (cf. Table\,\ref{deltatable}) and the CO data show no
significant variation of the index with velocity channel width
when averaged over all velocity bins (Figure~\ref{slopevel}a). The
indices stay relatively constant. Since the bulk of the molecular
gas traced by CO is even colder than cold \ion{H}{i} gas, these
results suggest a sequence of a reduced dependence of the spectral
index of the channel maps on the bin widths from the warm to the
cold ISM. The constancy of the average spectral index could also
be explained by optical depth effects. \citet{lp04} have shown
that absorption can lead to an effective slice broadening, which
leads in extreme cases to slice indices that become independent
from the actual channel width.

\begin{figure}
   \centering
   \includegraphics[width=\linewidth]{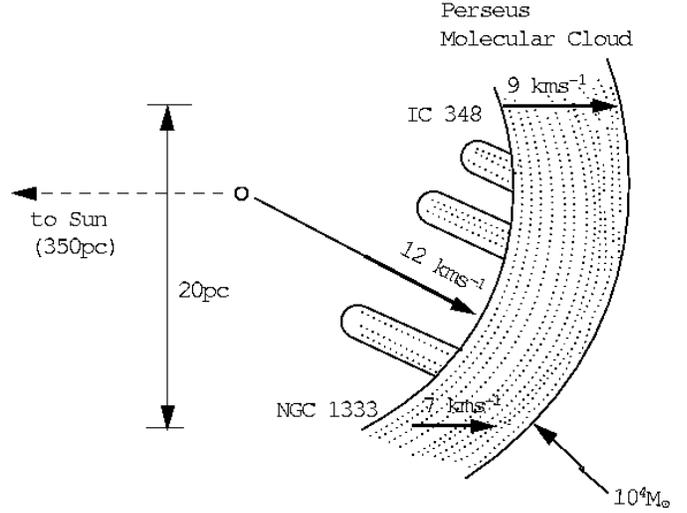}

   \caption{Sketch adopted from Fig.~3 of \citet{san74}
illustrating the spatial arrangement and motion of the Perseus
cloud complex. The gas is swept up by a shock expansion with
12~km\,s$^{-1}$. Due to the overall curvature, the line-of-sight
velocity is 9~km\,s$^{-1}$ for IC\,348, but only 7~km\,s$^{-1}$
for NGC1333. The diameter of the cloud is $\sim$ 20\,pc.
Pillar-like structures are left at lower velocities as remainders
of high-density regions which were not accelerated to the same
velocity.} \label{explode}
\end{figure}

Studying the spectral index of individual velocity bins of the CO
data across the line profile (Fig.\,\ref{deltachan}), we find that
the power-law indices increase with the velocity channel width in
the blue wing (Fig.\,\ref{slopevel}b) while staying rather
constant in the red wing.  No corresponding analysis was conducted
for the \ion{H}{i} data by \citet{dmsgg01}. One possibility to
explain the asymmetry between the blue and red wings might be a
shock expansion of the CO gas in Perseus. \citet{san74} found an
expanding shell of neutral hydrogen which was created by a
supernova in the Per\,OB2 association a few $10^6$ years ago. At
the location of the molecular cloud complex, the expansion is
directed away from the Sun. Most of the associated molecular gas
has been swept up by the shock, but pillar-like filaments have
been left at the backside of the shock. They are visible in the
channel maps (Fig. \ref{velchan}) and produce the velocity
dependence seen in the VCA of the blue line wings, i.e. the
increase of indices with size of the velocity bin because of a
gradual increase of large scale contributions across the line
wing. We illustrate this scenario in Fig.~\ref{explode}.

The quantitative results from the velocity channel analysis can be
interpreted in terms of the power spectrum of the velocity
structure. \citet{lp00} showed that the spectrum of velocity
slices as a function of the velocity channel width is determined
by the power spectral indices of the density structure $\beta$ and
the velocity structure $m$. They obtain different regimes for
shallow ($\beta < 3$) and steep ($\beta > 3$) density power
spectra:
\begin{center}
$
P(k) \propto \left\{%
\begin{array}{ll}
    k^\mathrm{-\beta+\frac{m-3}{2}}, & \hbox{thin slices,} \\
    k^\mathrm{-\beta}, & \hbox{thick slices,} \\
    k^\mathrm{-\beta}, & \hbox{very thick slices;} \\
\end{array}%
\right. (\beta < 3)$
\end{center}
\begin{center}
$
P(k) \propto \left\{%
\begin{array}{ll}
    k^\mathrm{-\frac{9-m}{2}}, & \hbox{thin slices,} \\
    k^\mathrm{-\frac{3+m}{2}}, & \hbox{thick slices,} \\
    k^\mathrm{-\beta}, & \hbox{very thick slices.} \\
\end{array}%
\right. (\beta > 3)$
\end{center}
Thin slices have a velocity width less than the local velocity
dispersion at the studied scale; thick slices have a width larger
than the velocity dispersion and very thick slices essentially
correspond to the integrated maps \citep{lp00}. We can assume that
a single channel of our data corresponds to thin slices as they
are much narrower than any observed line width. We use $\beta$
from the integrated maps and the index measured in the single
channels to derive $m$. From the indices obtained by averaging
over the full line profile (Fig.~\ref{slopevel}a), we obtain $m$
values of 3.9$\pm$1.6, 3.9$\pm$1.9 and 3.8$\pm$1.7 for $^{12}$CO
3--2 in IC\,348, NGC\,1333 and L\,1455, respectively; while $m$ is
3.9$\pm$2.0, 3.5$\pm$2.5 and 4.1$\pm$1.8 for $^{13}$CO 2--1 in the
same three regions. In the blue wings (Fig.~\ref{slopevel}b), we
obtain $m$ values of 3.2$\pm$1.6, 3.4$\pm$1.4 and 3.6$\pm$1.5 for
$^{12}$CO 3--2; while $m$ is 4.3$\pm$1.9, 4.3$\pm$1.7 and
4.1$\pm$1.5 for $^{13}$CO 2--1 in the three regions. These values
have large error bars, so that they are not directly suited to
discriminate between different turbulence models. At least, we
find that all values are consistent with Kolmogorov turbulence
that gives $m \sim 3.7$ \citep{kol41}.

\section{Summary}

\begin{enumerate}

\item We present KOSMA maps of the $^{13}$CO 2--1 and $^{12}$CO
  3--2 emission of the Perseus molecular cloud covering 7.1 deg$^{2}$. These data are
  combined with FCRAO maps of integrated $^{12}$CO and $^{13}$CO 1--0
  intensities and with a 2MASS map of optical extinctions.

\item To characterize the cloud density structure, we applied the
  $\Delta$-variance analysis to integrated intensity maps. The $\Delta$-variance
  spectra of the overall region follow a power law with an index of $\beta=2.9-3.0$
  for scales between 0.2 and 3\,pc. This agrees with results obtained by \citet{pad03a}
  studying structure functions of a $^{13}$CO 1--0 map of Perseus.

\item We also applied the $\Delta$-variance method to seven
  sub-regions of Perseus. The resulting power spectral indices vary significantly
  between the individual regions.  The active star-forming region NGC\,1333
  shows high spectral indices ($\beta=3.5-3.8$) while the dark
  cloud L\,1455 shows low indices of 2.9 in both transitions.

\item Additional information is obtained from the $\Delta$-variance
   spectra of individual velocity channel maps. They are very sensitive
   to optical depth effects, indicating self-absorption in the densest
   regions. The asymmetry of the channel map indices relative
   to the line centrum is a hint towards a peculiar velocity structure
   of the Perseus cloud complex.

\item When analyzing the spectral indices as a function of the
   velocity channel width we find almost constant indices when
   averaging over the total line profile. A continuous increase of
   the index with varying velocity channel width
   is, however, observed in the blue wings. This behavior can be explained by
   a shock running through the region creating a filamentary
   structure preferentially at low velocities.

\end{enumerate}

We find that the comparison of the structural properties for
entire surveys and sub-sets, as well as the velocity channel
analysis (VCA), provide additional, significant characteristics of
the ISM in observed CO spectral line maps. These quantities are
useful for a comparison of the structure observed in different
clouds, possibly providing a diagnostic tool to characterize the
star-formation activity and providing additional constraints for
numerical simulations of the ISM structure.

\begin{acknowledgements}

We are very grateful to helpful discussions with A. Lazarian and
D. Johnstone. We also thank Joao Alves for providing us the
2MASS extinction data.

\end{acknowledgements}

\bibliographystyle{aa}
\end{document}